\documentclass[aps,prl,reprint,superscriptaddress,twocolumn,showkeys,amsmath,amssymb,longbibliography,floatfix]{revtex4-2}
\usepackage[english]{babel}
\usepackage{amsmath,amssymb,bbm,mathrsfs,bm,braket,color,graphicx,comment,amsfonts,dsfont}
\usepackage[colorlinks,linkcolor=blue,citecolor=blue,urlcolor=blue]{hyperref}
\usepackage[mathscr]{euscript}
\usepackage{physics}
\usepackage{xcolor}
\usepackage[normalem]{ulem}
\usepackage{bm}
\usepackage{orcidlink}
\usepackage{multirow}
\usepackage{microtype}
\usepackage{bbold}

\usepackage{pifont}

\begin{document}

\title{
Family of Aperiodic Tilings with Tunable Quantum Geometric Tensor
}
\author{Hector Roche Carrasco\orcidlink{0009-0009-8824-7684}
}
\email{hector.roche@dipc.org}
\affiliation{Univ. Grenoble Alpes, CNRS, Grenoble INP, Institut Néel, 38000 Grenoble, France}
\affiliation{Donostia International Physics Center (DIPC),
Paseo Manuel de Lardiz\'{a}bal 4, 20018, Donostia-San Sebasti\'{a}n, Spain}
\author{Justin Schirmann\orcidlink{0009-0007-7030-0155}
}
\email{justin.schirmann@neel.cnrs.fr}
\affiliation{Univ. Grenoble Alpes, CNRS, Grenoble INP, Institut Néel, 38000 Grenoble, France}
\author{Aurelien Mordret\orcidlink{0000-0002-7998-5417}}
\email{aurmo@geus.dk}
\affiliation{Department of Geophysics and Sedimentary Basins, 
Geological Survey of Denmark and Greenland (GEUS), Øster Voldgade 10, 1350 Copenhagen K, Denmark}

\author{Adolfo G. Grushin\orcidlink{0000-0001-7678-7100}}
\email{adolfo.grushin@dipc.org}
\affiliation{Univ. Grenoble Alpes, CNRS, Grenoble INP, Institut Néel, 38000 Grenoble, France}
\affiliation{Donostia International Physics Center (DIPC),
Paseo Manuel de Lardiz\'{a}bal 4, 20018, Donostia-San Sebasti\'{a}n, Spain}
\affiliation{IKERBASQUE, Basque Foundation for Science, Maria Diaz de Haro 3, 48013 Bilbao, Spain}

\date{\today}

\begin{abstract}
    The strict geometric rules that define aperiodic tilings lead to the unique spectral and transport properties of quasicrystals, but also limit our ability to design them.
    In this Letter, we explore a novel example of a continuously tunable family of two-dimensional aperiodic tilings in which the underlying real-space geometry becomes a control knob of the wavefunction's quantum geometric tensor. The real-space geometry can be used to tune into topological phases occupying an expanded phase space compared to crystals, or into a disorder-driven topological Anderson insulator. The quantum metric can also be tuned continuously, opening new routes towards tunable single- and many-body physics in aperiodic solid-state and synthetic systems.
\end{abstract}

\maketitle

\textit{Introduction---}%
The electronic properties of quasicrystals 
can be strikingly different to those of crystals~\cite{Janot1992,Stadnik1999,Janssen2008}.
{For example, quasicrystals often exhibit rotational symmetries forbidden in crystals and wavefunctions which are neither localized nor fully extended~\cite{Janot1992,Stadnik1999,Janssen2008,Harper_1955,aubry1980analyticity}.}
However, the strict mathematical rules that determine the quasicrystalline 
atomic arrangements and the special wavefunctions they host also 
restrict the possibility of tuning them.
{
Different realizations of two-dimensional quasicrystals, such as the Penrose~\cite{Penrose74} and Ammann–Beenker~\cite{Shephard1992} tilings, cannot be smoothly deformed into one another because of their different symmetries. 
Nonetheless, smoothly connected families of quasiperiodic structures can be generated via deformed model sets~\cite{Welberry_2006_Deformed,baake2023dynamics}.
Continuously tuning the electronic properties of quasicrystals is also possible} by changing model parameters, 
as in crystals~\cite{Kraus:2012iqa,Tran:2015cj,Bandres:2016gx,Fuchs:2016hp,Huang2018,Huang:2018gu,Fuchs:2018dd,Loring2019,varjas_topological_2019,Chen2019,He2019,Duncan2020,Zilberberg:21,Hua2021,Jeon2022,Schirmann_2024}.

\begin{figure}[ht]
    \centering    \includegraphics[width=\linewidth]{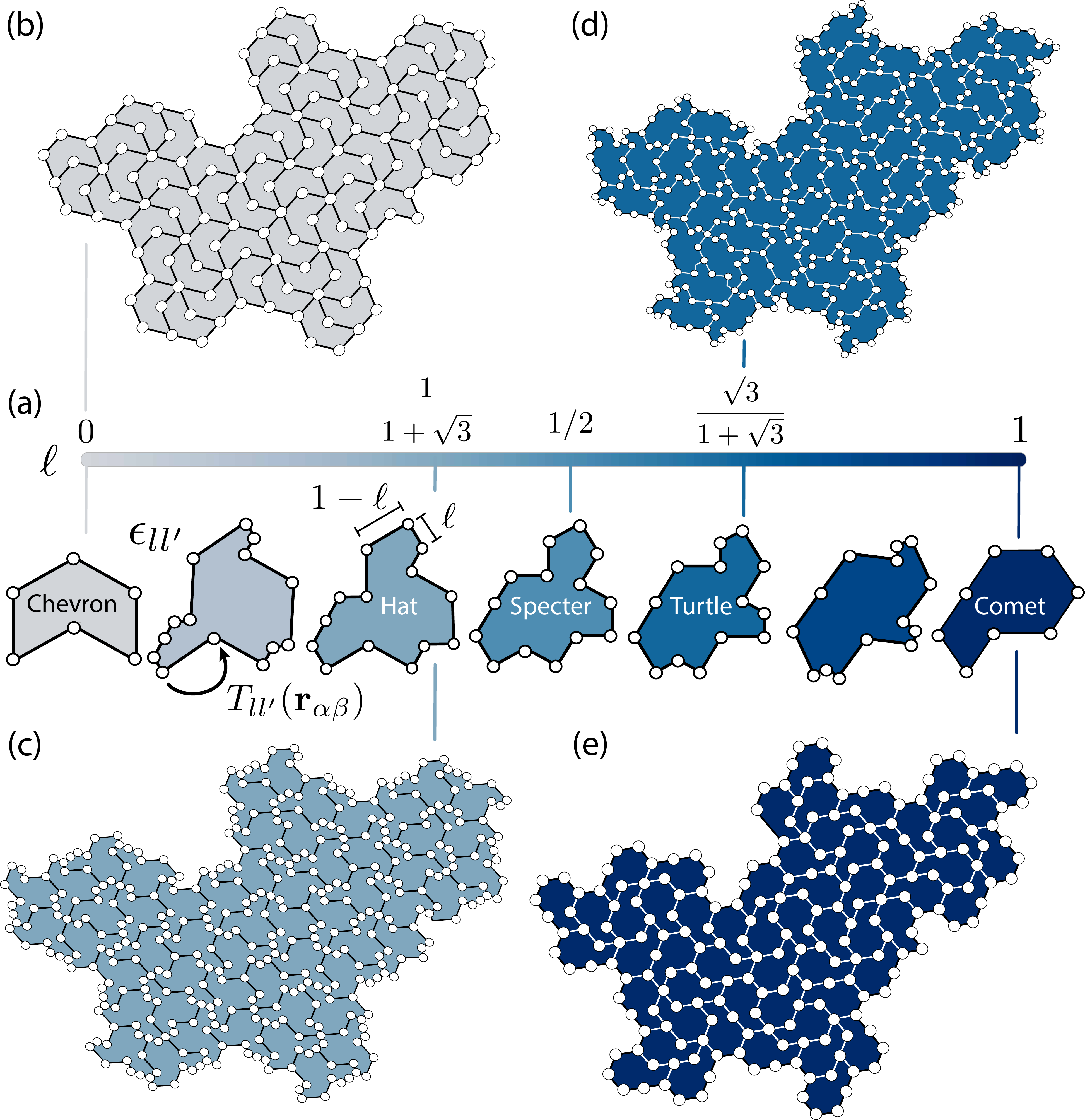}
\caption{\label{fig:fig1} 
Tiles and tilings for different values of the tile side length, defined by two side lengths, $\ell$ and $1-\ell$.
(a) shows a selection of tiles, including the Chevron, Hat, Turtle, and Comet. Their corresponding aperiodic tilings are shown in (b-e). For each $\ell$ we place two orbitals $l$ at every vertex, with onsite terms $\epsilon_{ll'}$. Pairs of sites $\{\alpha,\beta\}$ are connected by the space and orbital dependent hopping $T_{ll'}(\mathbf{r}_{\alpha\beta})$, see Eq.~\eqref{eq:hamiltonian}.}
\end{figure}
A striking example of continuously tunable electronic properties are moir\'{e} heterostructures—stacks of two-dimensional materials rotated relative to each other by a twist angle~\cite{Cao2018a,Cao2018b,Andrei2020,T_rm__2022}.
Tuning this angle alters the real-space geometry, which in turn modifies the two components of the wavefunction’s quantum geometric tensor, the quantum metric and Berry curvature~\cite{Fubini1904Metric,Study1905Metric,cheng2013quantumgeometric,yu2024quantumgeometryquantummaterials,LiuQuantumMetricReview,TormaWhereQG2023,verma2025quantumgeometryrevisitingelectronic}.
The external control over this tensor has catalyzed the discovery of a wide range of phenomena~\cite{Kennes2021review,T_rm__2022}, including superconductivity~\cite{Cao2018b,Yankowitz2019,Chen2019TTG,Lu2019}, anomalous and fractional Hall effects~\cite{Cai2023,Zeng2023,Park2023,Xu2023,Kang2024,Lu2024,Xie2024,Waters2024,Lu2024b}, Wigner crystallization~\cite{Regan2020,Xu2020,Li2021,Li2024}, quasicrystalline behavior~\cite{Ahn_2018,Uri2023,Lai2023,Hao2024,Guerrero2025}, and correlated states~\cite{Cao2018a,Choi2019,Sharpe2019,Polshyn2019,Codecido20219,Liu2020TBG,Shen2020,Chen2020TBG,Cao2020}.
However, these remarkable phenomena often occur at specific twist angles, and additionally suffer from twist-angle disorder~\cite{Li2010,Brihuega2012,Uri2020}.
Moreover, synthetic platforms like photonic crystals or ultracold atoms cannot be easily twisted.
The successes—but also the limitations—of moiré heterostructures therefore motivate the search for new ways to optimize the quantum geometric tensor in solid-state and synthetic matter via real-space geometry.

Here we show that a newly discovered family of aperiodic tilings~\cite{Smith2023a,Smith2023b} 
offers a real-space geometric pathway to tune the quantum geometric tensor.
{This 
family of aperiodic tilings is parametrically connected by a single real parameter $\ell$ 
offering a new method to geometrically tune electronic properties.} 
This parameter determines the position of all atomic sites while preserving the quasicrystalline long-range order.
Allowing the hoppings to depend on atomic positions, we show that changes in $\ell$ translate into changes of the quantum geometric tensor~\cite{Abouelkomsan2023,yu2024quantumgeometryquantummaterials,verma2025quantumgeometryrevisitingelectronic}, expressed in terms of a real-space Chern marker~\cite{Resta2011} and quantum metric~\cite{Marzari1997,MarsalQuantumMetric2024}. 
Changes in the real-space Chern marker induced by $\ell$ induce phase transitions in and out of a topological phase, which exists in an expanded parameter space compared to the same model in crystals.
The parameter $\ell$ also tunes the trace of the quantum metric \cite{Fubini1904Metric,Study1905Metric,cheng2013quantumgeometric,yu2024quantumgeometryquantummaterials,LiuQuantumMetricReview}, which contributes to the superfluid stiffness and kinetic inductance~\cite{JulkuSuperfluidLiebFlatBand2016,Tovmasyan2016FlatBandHubbard,XieSuperfluidWeight2020,T_rm__2022,Bernervig2022Superfluid,peotta2023quantumgeometrysuperfluiditysuperconductivity}, enters optical sum rules~\cite{qiu2024quantumgeometryprobedchiral,Onishi_2024,chiu2025topologicalsignaturesopticalbound}, and determines localization~\cite{chau2024disorderinduceddelocalizationflatbandsystems} and entanglement properties~\cite{Nisarga2024Arealaw}.
Hence, changing the real-space geometry using $\ell$ allows one to tune the quantum geometric tensor.

\begin{figure*}
    \centering    \includegraphics[width=\linewidth]{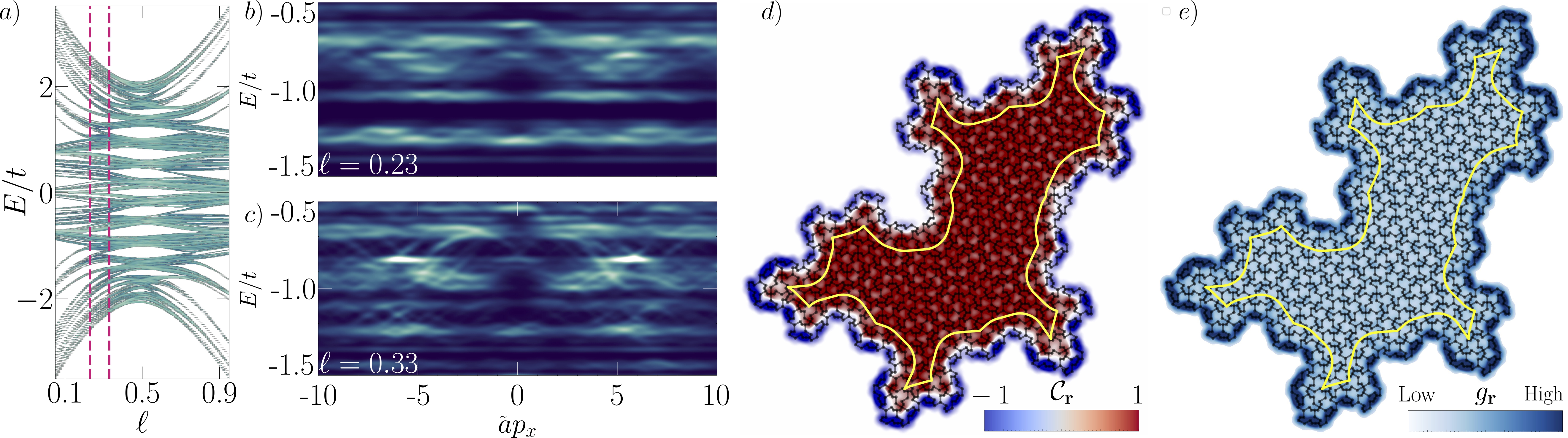}
\caption{\label{fig:spectral_function}
Spectral, topological, and geometric properties of the model for different values of $\ell$. (a) Energy spectrum of the bulk Hamiltonian as a function of $\ell$. 
We obtain the bulk spectrum by excluding states whose weight inside the yellow contour in (d) and (e) is smaller than their weight outside it. Magenta dashed lines indicate the $\ell$ values used in panels (b) and (c). The yellow dashed line indicates the Fermi energy $E_F=-1.15t$.
(b) Spectral function $\mathcal{A}(E,{\bf p})$ for $\ell=0.23$, {showing a full gap near 1/4 filling, attained} near the energy indicated by the white dashed line. (c) Spectral function $\mathcal{A}(E,{\bf p})$ for $\ell=1/3$, showing the presence of edge states within bulk gaps. (d) Local Chern marker $\mathcal{C}_{{\bf r}_i}$ Eq.~\eqref{eq:LCM}, evaluated at the energy indicated by the white dashed line in (c) for $\ell=1/3$. The yellow solid line outlines the bulk region used to perform the averaging. In this region, the average value of the local Chern marker is $0.98$. (e) Local quantum metric $g_{\bf r}$, Eq.~\eqref{eq:QM_marker}, computed at the $E_F$ indicated by the dashed line in (c) for $\ell=1/3$. We observe that the local quantum metric takes large values at the edges, indicating delocalized edge states.  In all the panels the system has $2530$ sites and $M/t=-2.7$. In panels (b) and (c) we used the kernel polynomial method~\cite{Wei_KPM_2006} to compute the spectral function with 512 moments.}
\end{figure*}

\textit{Geometric control of topology---}%
Fig.~\ref{fig:fig1}a shows the Hat family of aperiodic tilings, parametrized by a real number $\ell\in \left[0,1\right]$.
It parametrizes the two independent lengths of parallel sides, $a=\ell$ and $b=1-\ell$.
By tuning $\ell$, one can continuously interpolate between different tiles in the family.
The Hat tiling [$\ell=1/(1+\sqrt{3})$] made headlines as the first aperiodic monotile~\cite{Smith2023a,Smith2023b}. 
Other named tiles are the \textit{Specter} [$\ell=1/2$], the \textit{Turtle} [$\ell=\sqrt{3}/(1+\sqrt{3})$], the \textit{Chevron} [$\ell=0$], and the \textit{Comet} [$\ell=1$] which sets our unit of length.
A previous Letter has shown that a non-bipartite version of the Hat tiling shares similar electron properties with those of graphene~\cite{Schirmann_2024} when all the hoppings are set to be equal{, inherited from its sixfold rotational symmetry~\cite{Socolar_2023_Quasicrystalline}}.
Here, we focus on the bipartite version of these tilings, and describe their construction in 
the Supplemental Material (SM)~\cite{SuppMat}, by adapting the code published along with Ref.~\cite{AGG_Seismic_2025}.

    \begin{figure}[h!]
        \centering    \includegraphics[width=\linewidth]{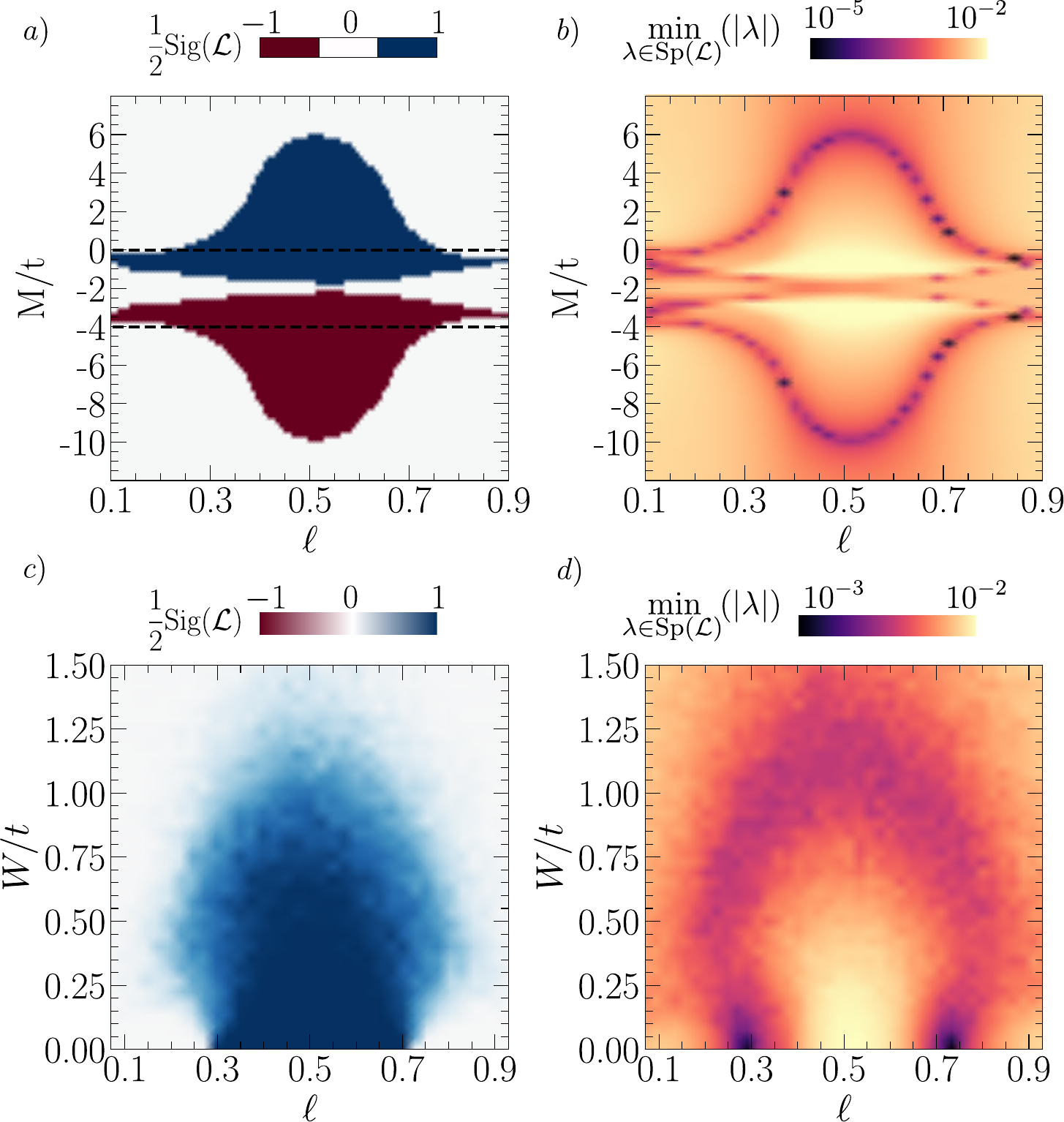}
    \caption{\label{fig:phase_diag} 
    Topological phase diagram. (a) Bulk half-signature of the spectral localizer at $E_0=E_F$ as a function of both the geometrical parameter $\ell$ and the value of the onsite term $M/t$. 
    $E_F$ is fixed at $3/4$ filling out of a total of $N_\mathrm{sites}\times N_{\mathrm{orbs}}=2530\times 2$ states. We observe topological phases for a wider parameter region compared to the square-lattice crystalline model, whose phase boundaries are delimited by the black dashed lines. (b) Localizer gap corresponding to (a). The gap in the spectrum of the spectral localizer closes at the boundary between the topological and trivial phases of panel (a). (c) Bulk half-signature of the spectral localizer averaged over 100 disorder realizations at $E_0=E_F$ and $M/t = 0.5$, as a function of $\ell$ and onsite disorder strength $W/t$. We observe two types of behavior as we increase $W/t$ for fixed $\ell$: either the clean system is topological and transitions into a trivial Anderson insulator (e.g. at $\ell=0.5$), or a trivial clean system transitions into a topological phase, to later become a trivial Anderson insulator (e.g. at $\ell=0.8$ or $\ell=0.2$). (d) The localizer gap vanishes between the topological and trivial Anderson insulators in (c). In all panels, the system has $N_\mathrm{sites}=2530$ sites and $\kappa = 0.01$.
    {Panels (b) and (d) are plotted using a logarithmic colormap.}}
    \end{figure}
To exemplify how $\ell$ can control the electronic quantum geometry we consider a non-crystalline generalization of the Qi-Wu-Zhang Chern insulator model~\cite{QWZ_QSH_2006} on the vertices of the tiling, see Fig.~\ref{fig:fig1}a-e,
that reads \cite{bernevig_quantum_2006,Bernevig2013_TI_TS,agarwala_topological_2017,Liu_ChernHyperbolic_2022}:
\begin{align}
    \label{eq:hamiltonian}
    \mathcal{H} =& \sum_{\left\langle\alpha,\beta \right\rangle}\sum_{l,l' \in\left\{1,2\right\}} 
    T_{ll'}^{\phantom{}}(r_{\alpha\beta}^{\phantom{}},\theta_{\alpha\beta}^{\phantom{}}) c^\dagger_{\alpha,l} c_{\beta,l'}^{\phantom{}} \nonumber\\
    &+\sum_{\alpha}\sum_{l,l' \in\left\{1,2\right\}} \epsilon_{ll'}^{\phantom{}}c_{\alpha,l}^{\dagger}c_{\alpha,l'}^{\phantom{}}.
\end{align}
The hopping amplitude between two orbitals $l$ and $l'$ at sites $\alpha$ and $\beta$ sharing an edge, $T_{ll'}(r_{\alpha\beta},\theta_{\alpha\beta})$, depends on the inter-atomic separation $r_{\alpha\beta}=|{\bf r_\alpha}-{\bf r_\beta}|$ 
and the polar angle $\theta_{\alpha\beta}$.
We choose the hopping to be
$T(r_{\alpha\beta},\theta_{\alpha\beta}) = h(\theta_{\alpha\beta})f(r_{\alpha\beta})$, 
with $f(r_{\alpha\beta}) = \exp(1-r_{\alpha\beta}/\bar{a})$ and
$h(\theta_{\alpha\beta})= -(1/2)\sigma_z-(i t/2)\left[\cos(\theta_{\alpha\beta})\sigma_x +\sin(\theta_{\alpha\beta})\sigma_y\right]$, and the onsite energy as
$\epsilon = (2+M)\sigma_z$. Here
$\bar{a}$ denotes the average bond-length and $(\sigma_x,\sigma_y,\sigma_z)$ are the Pauli matrices acting in orbital space.
If we would choose $\alpha,\beta$ to lie on a square lattice we would recover the crystalline Chern insulator of Ref.~\cite{QWZ_QSH_2006}.

The corresponding bulk energy spectrum is shown in Fig.~\ref{fig:spectral_function}a as a function of $\ell$. 
It is symmetric with respect to zero energy as a consequence of a chiral symmetry~\cite{agarwala_topological_2017} that places the model in class $D$ of the Altland-Zirnbauer classification~\cite{altland97}.
This feature is not essential to our findings.
At the extreme values $\ell\approx 0,1$ the spectrum is gapped and we expect the bands to be topologically trivial.
The reason is that at these values of $\ell$ many sites come together in real space, forming clusters of sites (see Fig.~\ref{fig:fig1}a and Supplemental Video). 
The exponential form of $f(r_{\alpha\beta})$ leads to a strong suppression of the hoppings between clusters, exponentially localizing the wavefunctions to these clusters.
Since exponentially localized wavefunctions are topologically trivial, so are the gaps~\cite{Panati2007,Brouder_LocalizedWannier_2007,Soluyanov2011}.

\begin{figure}
    \centering    \includegraphics[width=\linewidth]{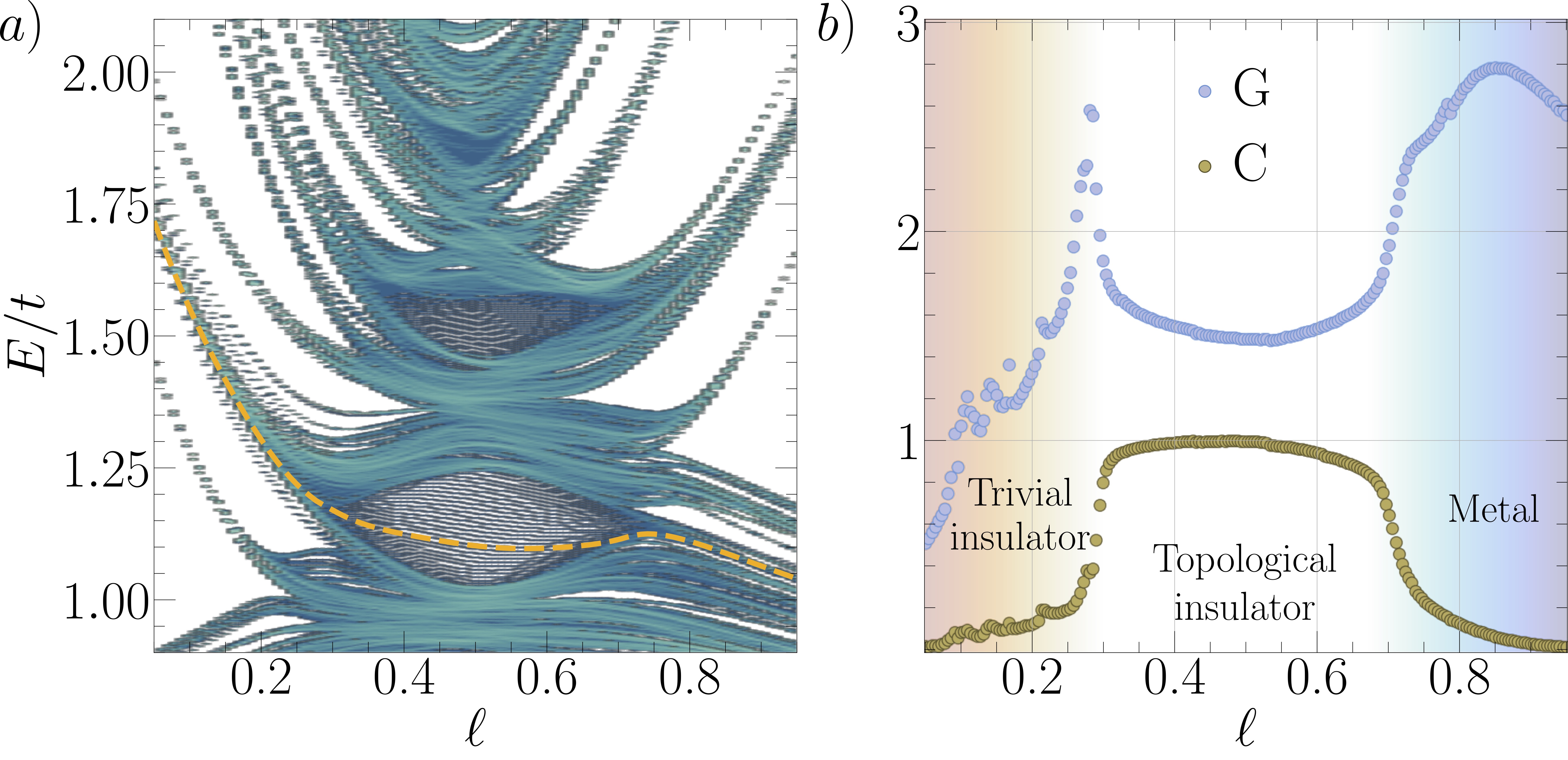}
\caption{\label{fig:flatness_QM} 
(a) Energy spectrum as a function of $\ell$, showing a zoomed-in version {of Fig.~\ref{fig:spectral_function}a} including bulk and edge states. $3/4$ filling is shown as a yellow dashed line in (a). (b) Bulk trace of the local quantum metric $G$ and local Chern marker $C$ for this filling as a function of $\ell$. 
At small $\ell$ the sites cluster and localize the wavefunctions for this filling, leading to a small $G$ with $C\approx 0$. At intermediate $\ell \approx 0.5$, $G\ge C \approx 1$, satisfying the trace inequality. For large $\ell \gtrsim 0.7$ this filling sits in the middle of a band of delocalized states with increased $G$.  In both panels, the system has $N_\mathrm{sites}=2530$ and $M/t=-2.7$.
}
\end{figure}

This expectation is confirmed by computing the bulk average of the local Chern marker \cite{bianco11}
\begin{equation}
\label{eq:LCM}
C_{\textbf{r}_\alpha}=2\pi \text {Im}\sum_{l}\bra{\textbf{r}_\alpha,l}
\hat{P}\left[\hat{Q}\hat{X},\hat{P}\hat{Y}\right]\hat{P}\ket{\textbf{r}_\alpha,l},
\end{equation}
where $\hat{P}=\sum_{E<E_F} \ket{E}\bra{E}$ is the projector onto occupied eigenstates $\ket{E}$ below the Fermi energy $E_F$, $\hat{Q}=1-\hat{P}$, and $\hat{X}$ and $\hat{Y}$ the position operators in the plane. 
Its average over a bulk real-space region of area $\mathcal{A}_b$ (see 
SM~\cite{SuppMat}), $C = (1/\mathcal{A}_b)\sum_{\alpha\in\text{Bulk}}C_{\textbf{r}_\alpha}$,  determines the bulk Chern number, $C$, which is approximately quantized if the occupied states are topological~\cite{Resta2011}.
We find that the average of the local Chern marker is {close to} zero ({see Fig.~\ref{fig:flatness_QM}}) when we place $E_F$ within any of the gaps seen at $\ell\approx0,1$, as advertised.
As we tune $\ell$, the angle- and position-dependence of the hoppings in Eq.~\eqref{eq:hamiltonian} allows the spectrum to evolve, see Fig.~\ref{fig:spectral_function}a.
At $\ell\approx 0.28$ we observe {gap closings across the whole energy range, suggesting potential} topological phase transitions.

Topological transitions would imply the appearance of in-gap spectral weight corresponding to topological edge states in the total momentum-resolved density of states, or spectral function.
We define the spectral function by projecting the real-space Hamiltonian $\mathcal{H}$ onto plane-wave states, $A(E,\mathbf{p})=\sum_{l}\bra{\textbf{p},l}\delta (\mathcal{H}-E)\ket{\textbf{p},l}$, where $\ket{\textbf{p},l}=\sum_{\alpha}e^{i\mathbf{p}\mathbf{r}_\alpha}\ket{\textbf{r}_\alpha,l}$, 
see 
SM~\cite{SuppMat}.
This function is measurable in angle-resolved photoemission spectroscopy even without translational invariance~\cite{Marsal2022,corbae2023,Ciocys2023,Rotenberg2000,Rotemberg2004,Rogalev2015}.
Note that, unlike the crystalline case, the components of ${\bf p}$ are not restricted to the interval $\left[0,2\pi\right]$.
Comparing Figs.~\ref{fig:spectral_function}b and c, at either side of the gap closing (with $\ell=0.23$ and $\ell=0.33$, respectively) we see the appearance of in-gap spectral weight, as expected for a topological transition (see also Supplemental Video).

We confirm the topological nature of the in-gap states at $\ell=0.33$ by computing the local Chern marker at $E_F=-1.15t${, corresponding to 1/4 filling,} see Fig.~\ref{fig:spectral_function}d. 
For these parameters its bulk average is $C \approx 0.98$, confirming the topological phase. 
These results show that we can tune the topological phase by varying $\ell$ 
while keeping the filling fixed.
For example, at $E_F =-1.15t$ {(yellow dashed line in Fig.~\ref{fig:spectral_function}a), 
by varying $\ell$, a state that is initially trivial becomes
topological.}

We now ask, how far do the topological properties survive in parameter space,
compared to the crystalline lattice model with the same Hamiltonian parameters?
{To answer this question it is convenient to use the spectral localizer index, which is
obtained from the spectral localizer}~\cite{LORING2015,loring2019guidebottindexlocalizer,cerjan_local_2022,Cerjan_2024_Tutorial}
\begin{equation}
    \mathcal{L} = (\hat{\mathcal{H}}-E_0\mathbb{1}) \sigma_z + \kappa\left[(\hat{X}-x_0\mathbb{1})\sigma_x+ (\hat{Y}-y_0\mathbb{1})\sigma_y\right],
\end{equation}
where $E_0$ is the reference energy at which we want to probe topology and $x_0$ and $y_0$ are chosen in the bulk of the system~\footnote{In practice, we choose the coordinates $x_0$ and $y_0$ by computing the geometric mean of the vertex coordinates.}. The scalar $\kappa$ is fixed to compensate the differences in scales between $\hat{\mathcal{H}}$ and the position operators $\hat{X}$, and $\hat{Y}$,  see 
SM~\cite{SuppMat}.
The half-signature of the localizer, $\frac 12 \text{Sig}(\mathcal{L})$, defined as the half-difference between the number of positive and negative eigenvalues, is an integer that equals the bulk Chern number~\cite{LORING2015,loring2019guidebottindexlocalizer,cerjan_local_2022,Cerjan_2024_Tutorial}.

{Although the local Chern marker and the spectral localizer provide equivalent topological information~\cite{jezequel2025explicitequivalencespectrallocalizer}, the spectral localizer additionally quantifies the robustness of the topological phase. In particular, its minimal eigenvalue, known as the localizer gap, serves as a measure of phase stability, since a topological phase transition can only occur when this gap closes~\cite{Cerjan_2024_Tutorial,SuppMat}. Because the localizer index does not require calculating projectors, topological phase diagrams can be computed numerically more efficiently than the local Chern marker average~\cite{Cerjan_2024_Tutorial}.}

As a reference, when applied on a square lattice, the model defined in Eq.~\eqref{eq:hamiltonian} only has topological phases when {$-4<M/t<0$}.
In contrast, we see in Fig.~\ref{fig:phase_diag}a that the phase diagram of the Hat family as a function of $\ell$ and $M/t$ extends further.
The topological bands survive 
for much larger values of $M/t$ compared to the square lattice model, delimited by the dashed lines in Fig.~\ref{fig:phase_diag}a, especially around $\ell \approx 0.5$
~\footnote{The edge statistics explain why topological regions widen around $\ell=0.5$, and why the phase diagram is asymmetric about this point. For generic $\ell$ there are $N_{s}$ short and $N_{l}$ long bonds, satisfying $N_{s}>N_{l}$ for $\ell<0.5$, $N_{s}<N_{l}$ for $\ell>0.5$, and $N_{\rm s}=N_{l}$ for $\ell=0.5$. Since the average spacing $\bar a$ in $f(r_{\alpha\beta})$ increases with $\ell$, this imbalance yields an asymmetric hopping distribution. At $\ell=0.5$ all hoppings are equal, minimizing detrimental bond disorder~\cite{SongJ2012,Lv2013} and leaving only connectivity disorder, as in Weaire–Thorpe models~\cite{weaire_electronic_1971}. Away from $\ell=0.5$, the two competing hopping strengths associated to $N_{s}$ and $N_{l}$ drive the system toward trivial, Wannierizable dimerized limits at $\ell\to0,1$, leaving $\ell=0.5$ as the optimal value.}.
Consistently, we see that the localizer gap, shown in Fig.~\ref{fig:phase_diag}b, closes at the phase boundaries.
Given that $\ell$ serves as a tuning parameter for inducing topological phase transitions, one might ask whether it can also be used to control a {disorder-induced topological phase} transition. To address this question, we introduce a random onsite potential
\begin{equation}
V = \sum_{\alpha}\sum_{ll'} V_{\alpha,ll'}^{\phantom{}}c_{\alpha,l}^{\dagger}c_{\alpha,l'}^{\phantom{}},
\end{equation}
where $V_{\alpha,ll'}^{\phantom{}} = V_\alpha \delta_{ll'}$ represents a site-dependent potential drawn from a uniform distribution in the range $[-W/2,W/2]$, where $W$ controls the disorder strength. 
{We observe in Fig.~\ref{fig:phase_diag}c that the system remains topological for disorder strengths up to $W/t \leq 1$, before undergoing an Anderson transition, signaled by the closing of the average localizer gap shown in Fig.~\ref{fig:phase_diag}d. As in crystals, for certain values of $\ell$ (e.g., $\ell=0.8$), the system, initially trivial in the clean limit, enters a topological phase as $W$ increases~\footnote{This phase
is sometimes referred to as topological Anderson insulator phase even though it is connected to the disorder-free limit}, indicating that disorder can enlarge the topological region in parameter space~\cite{Li2009AndersonInsulator,groth_theory_2009,khudaiberdiev2024twodimensionaltopologicalandersoninsulator}. Interestingly, $\ell$ serves as an additional tuning knob that enables a topological phase transition at fixed disorder strength, a mechanism specific to this tiling family.}

\textit{Geometric control of the quantum metric---}%
The local Chern marker is the imaginary part of a more general tensor called the quantum geometric tensor
\cite{Fubini1904Metric,Study1905Metric,cheng2013quantumgeometric,yu2024quantumgeometryquantummaterials,LiuQuantumMetricReview,TormaWhereQG2023,verma2025quantumgeometryrevisitingelectronic}
, which can also be controlled by tuning $\ell$.
Its real part is known as the quantum metric~\cite{TormaWhereQG2023,yu2024quantumgeometryquantummaterials,verma2025quantumgeometryrevisitingelectronic}, and is proportional to the real-space spread of occupied states.
By expressing it as a local real-space quantity 
(see 
SM~\cite{SuppMat} and Refs.~\cite{Marzari1997,MarsalQuantumMetric2024})
\begin{equation}
    \label{eq:QM_marker}
    g_{\bf{r}_{\alpha}} = 2\pi\mathrm{Re}\sum_{\mu = \left\{x,y\right\}}\sum_l \langle {\bf r}_{\alpha},l\lvert \hat{P}\hat{r}_{\mu}\hat{Q}\hat{ r}_{\mu}\hat{P}\rvert {\bf r}_{\alpha},l\rangle,
\end{equation}
where $\hat{r}_{x} = \hat{X}$ and $\hat{r}_y = \hat{Y}$,
we can use it to observe edge-state signatures, as shown in Fig.~\ref{fig:spectral_function}e for the topological state at $\ell=0.33$ at $E_F=-1.15t${, approximately at 1/4 filling}. 
We see $g_{\bf{r}_{\alpha}}$ reaches its maximum values at the edges of the system, as expected for delocalized topological edge states. {In contrast, for the trivial phase $g_{\bf{r}_{\alpha}}$ is peaked around clusters of sites, consistent with a trivial phase, see SM~\cite{SuppMat}.
}

As with the bulk local Chern marker, the bulk quantum metric, is also tunable as a function of the parameter $\ell$ at a given filling. We define the bulk quantum metric by averaging Eq.~\eqref{eq:QM_marker} over a real-space region of area $\mathcal{A}_b$ as $G = (1/\mathcal{A}_b)\sum_{\alpha\in\text{Bulk}}g_{{\bf r}_\alpha}$, see
SM~\cite{SuppMat}.
Tuning $G$ is desirable in several contexts.
For instance, $G$ is known to be bounded from below by the Chern number, $G \ge C$, an inequality known as the trace inequality~\cite{Souza2000,Parameswaran2013,Roy2014,Claassen2015,Ledwith2020,Wang2021}.
When a band saturates this inequality it is termed vortexable or ideal~\cite{Ledwith2020,Wang2021}, as it is a favorable starting point to realize fractional Chern insulator states~\cite{Liu_FCI_Flatband_2012,BERGHOLTZ_Flatband_FracChern,AbouelkomsanMoireFlatBandChern}.
The quantum metric enters the superfluid stiffness~\cite{JulkuSuperfluidLiebFlatBand2016,Tovmasyan2016FlatBandHubbard,XieSuperfluidWeight2020,T_rm__2022,Bernervig2022Superfluid,peotta2023quantumgeometrysuperfluiditysuperconductivity} and optical sum rules bounding topological gaps~\cite{Souza2000,qiu2024quantumgeometryprobedchiral,Onishi_2024,chiu2025topologicalsignaturesopticalbound}.
The quantum metric depends on the real-space position, or embedding, of the orbitals~\cite{Bena_2009,Fruchart_2014,Dobard2015,Lim2015,Cook2017,Simon2020,TormaWhereQG2023,Huhtinen2022}, controlled by $\ell$ in our system.
In crystals, the embedding-dependence can be exploited to optimize responses such as the photovoltaic efficiency~\cite{Cook2017}.
Fig.~\ref{fig:flatness_QM} shows the tunability of the quantum geometric tensor by plotting the trace of the local Chern marker $C$ and the local quantum metric $G$ as a function of $\ell$.
For 3/4 filling (dashed yellow line in Fig.~\ref{fig:flatness_QM}a) $G$ decreases as $\ell$ decreases with $C\approx 0$, see Fig.~\ref{fig:flatness_QM}b.
This is consistent with the fact that for these parameters the bands are narrow, topologically trivial, and states are localized due to the clustering of sites (see Supplemental Video).
In contrast, at intermediate values of $\ell \approx 0.5$ the local Chern marker is approximately quantized to $C \approx 1$. 
The quantum metric decreases but does not saturate the trace inequality $G \ge C$, as seen in other Chern insulator models~\cite{Komissarov2024,Romeral2025}.
At large $\ell \gtrsim 0.7$, $C$ decreases to zero while $G$ increases.
This indicates that the states at this filling are more delocalized, consistent with the fact that $E_F$ falls within a trivial band with a relatively large spread in energy, see Fig.~\ref{fig:flatness_QM}a.
{Saturating the trace inequality in the topological region may be possible by optimizing within the large parameter space of fillings and model parameters, an interesting direction for future work.}

\textit{Conclusions---}%
In this work we have used the real-space geometric parameter $\ell$, which continuously connects members of the Hat family of aperiodic tilings, to tune the quantum geometric tensor. 
The parametrization in terms of $\ell$ is unique to this family, offering a new knob to control transport properties of aperiodic tilings.
By allowing the hopping amplitudes to depend on the relative positions of the atoms, we have shown that $\ell$ controls the transition between trivial and topological phases, also in the presence of disorder. The parameter space supporting topological phases is broader than that of the square-lattice counterpart, highlighting the enhanced tunability offered by these tilings.

We have shown how $\ell$ tunes the quantum metric, which underlies key physical properties such as the superfluid stiffness~\cite{JulkuSuperfluidLiebFlatBand2016,Tovmasyan2016FlatBandHubbard,XieSuperfluidWeight2020,Bernervig2022Superfluid,T_rm__2022,Thumin2023,peotta2023quantumgeometrysuperfluiditysuperconductivity} and the optical weight \cite{Verma2021OpticalWeightPhaseStiff,Onishi_2024,ghosh2024probingquantumgeometryoptical,Janjowski2025OpticalWeight}. The ability to tune the quantum metric suggests avenues to engineer properties of unconventional superconductors~\cite{shavit2024quantum}, photonic systems~\cite{Bleu_2018,petrides2023probingelectromagneticnonreciprocityquantum}, and correlated electron systems~\cite{Hassan_2018,chen2022measurementinteractiondressedberrycurvature,Kashihara_2023,Hu2025CoherenceLength}.

The continuous parametrization in terms of $\ell$ can be exploited in different physical contexts.
The systems we consider here could be realized in electronic systems by lithographic patterning of solid-state systems~\cite{Lassaline2021}, depositing CO molecules on metals~\cite{Gomes2012,Kempkes2019}, or by engineering 2D Shiba states with magnetic ad-atoms on superconductors~\cite{nadj2013proposal,Schneider2022,Soldini2023}. Moir\'{e} heterostructures have also been used to realize unconventional tessellations~\cite{Park2025}. 
Choosing different $\ell$ provides a new tuning knob to design topology and geometry in photonic metamaterials~\cite{Notomi2004,Levi2011, Kraus:2012iqa,Bandres2016,Ozawa2019_rev_photonics,moritake2025chiraldiffractionaperiodicmonotile},
polaritonic systems~\cite{Tanese2014}, electrical circuits~\cite{Stegmaier2023}, microwave networks~\cite{Vignolo2016}, acoustic~\cite{Chen2020} and mechanical~\cite{He2016,Ding2019,Wang2020b} metamaterials.  More broadly, the tunability of this family of aperiodic tilings has already proven useful in enhancing earthquake detection~\cite{AGG_Seismic_2025} and could benefit other areas such as the study of spin models~\cite{okabe2024ising}, elasticity~\cite{rieger2024macroscopic,naji2024effective}, and fluids~\cite{plawsky2025transport} on aperiodic tilings.

\textit{Acknowledgments} ---
We thank  E. Ben Achour, P. D'Ornellas, F. Flicker, L. Gómez Paz, Q. Marsal, S. Pachhal, M. Thumin, D. Varjas, and M. Yoshii for useful discussions and suggestions. We are grateful to  P. D'Ornellas and S. Roche for their critical reading of the manuscript. Technical and human support provided by the DIPC Supercomputing Center is gratefully acknowledged.
J. S. is supported by the program QuanTEdu-France n° ANR-22-CMAS-0001 France 2030.
A. G. G. is supported by the European Research Council (ERC) Consolidator grant under Grant Agreement No. 101042707 (TOPOMORPH). 

{The data and codes that support the findings of this article are openly available at~\cite{rochecarrasco_zenodo_2025}}.

\clearpage
\newpage
\onecolumngrid

\setcounter{secnumdepth}{5}
\renewcommand{\theparagraph}{\bf \thesubsubsection.\arabic{paragraph}}

\renewcommand{\thefigure}{S\arabic{figure}}
\setcounter{figure}{0} 

\appendix
\begin{center}
{\bf Supplementary Material for ``Family of Aperiodic Tilings with Tunable Quantum Geometric Tensor''}
\end{center}
\section{Generating aperiodic tilings from inflation rules}
\label{App:InflationRules}
\begin{figure}
    \centering    \includegraphics[width=\linewidth]{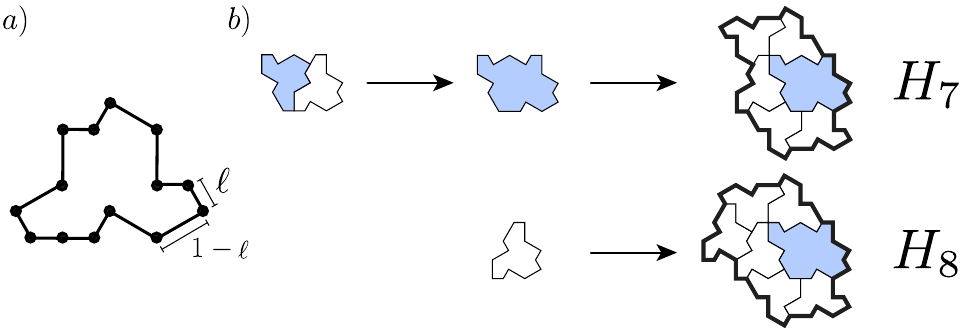}
\caption{\label{fig:Hat} 
Tile and inflation rule.
(a) A tile is a fourteen-sided polygon with edge lengths parameterized by a geometric parameter $\ell$. In this example, the value of $\ell$ is taken to be $1/(1+\sqrt{3})$.
(b) To construct the tiling, we use clusters of tiles. We begin by defining a two-tile compound consisting of one tile (in white) and its mirror image (in pale blue). We then establish a substitution rule for both the single tile and the two-tile compound using the \textit{H7} and \textit{H8} clusters.}
\end{figure}

In this Appendix, we describe the inflation procedure introduced in~\cite{Smith2023a} and employed in~\cite{weblink} to generate the aperiodic tilings we use in the main text. As illustrated in Fig.~\ref{fig:Hat}, this procedure involves the successive replacement of one set of tiles with another made of a larger amount of tiles. In this work, we adapt the code used in Ref.~\cite{AGG_Seismic_2025} to generate the tilings presented in Figs.~\ref{fig:fig1} and~\ref{fig:spectral_function}. We begin by examining a single tile, which is a 14-sided polygon whose edges can be grouped into pairs of equal-length, parallel segments, as shown in Fig.~\ref{fig:Hat}a. The first step is to combine one tile with its mirror image, forming a two-tile compound. Following~\cite{Smith2023a}, this two-tile compound can then be combined with six additional tiles to create a cluster, referred to as \textit{H8}, or with five other tiles to form the \textit{H7} cluster. The inflation process proceeds as follows: each individual tile is replaced by an \textit{H8} cluster, while each two-tile compound is replaced by an \textit{H7} cluster. This process can be iterated indefinitely to generate arbitrarily large lattices. The number of tiles after the $n$-th iteration is given by the following recursion relation
\begin{align}
    \left(H_7\right)_{n+1} &= 5 \left(H_8\right)_{n}+\left(H_7\right)_{n},\\\left(H_8\right)_{n+1} &= 6\left(H_8\right)_{n}+\left(H_7\right)_{n},
\end{align}
where 
$\left(H_7\right)_{n}$
(resp. $\left(H_8\right)_{n}$) denotes the number of tiles present in the \textit{H7} (resp. \textit{H8}) cluster after the $n$-th iteration of the inflation procedure. In the main text we present results for the third iteration, which has 2530 sites. We have checked that our results are well converged with system size.

\section{Bulk and surface spectral function.}
\label{App:SpectralFunction}
In this Appendix we describe explicitly the procedure followed to compute the spectral function shown in Fig.~\ref{fig:spectral_function}. We first introduce the real space Green's function. For a system composed of $N_{\text{sites}}$ atoms and $N_{\text{orbs}}$ internal degrees of freedom, the real-space Green's function is an $N_{\text{sites}}N_{\text{orbs}}\times N_{\text{sites}}N_{\text{orbs}}$ matrix which reads
\begin{equation}
    \mathcal{G}(E) = \lim_{\eta\rightarrow0} \left(\mathcal{H}-(E-i\eta)\mathbf{1}_{N_{\text{sites}}N _{\text{orbs}}}\right)^{-1},
\end{equation}
where $\mathcal{H}$ is the real-space Hamiltonian and $E$ is the energy at which the Green's function is computed. In practice, $\eta$ is taken to be finite. The value of $\eta$ is related to the number of moments in the expansion in terms of {Chebychev} polynomials used in the kernel polynomial method~\cite{Wei_KPM_2006}. 
We now introduce the tight-binding basis $\lvert {\bf{r}_{\alpha}},l\rangle $ that span the Hilbert-space. Each  $\lvert {\bf{r}_{\alpha}},l\rangle $ is localized on a site $\alpha$ and carries an internal orbital degree of freedom labeled $l$.
These states are represented as $N_{\text{Sites}}\times N_{\text{Orbs}}$ vectors and satisfy the following relation
\begin{align}
\label{eq:OrthoTBStates}
    \langle{\bf r}_{\alpha},l'\lvert {\bf{r}}_{\beta},l \rangle = \delta_{\alpha\beta}\delta_{ll'}.
\end{align}
Using these, we introduce the normalized plane-wave basis set
\begin{align}
\label{eq:PlaneWaves}
    \lvert {\bf{p}},l \rangle = \frac{1}{\sqrt{N_{\text{sites}}}}\sum_{\alpha}\exp\left(i {\bf p}\cdot {\bf r}_{\alpha}\right)\lvert {\bf r}_{\alpha},l\rangle.
\end{align}
One can now compute the total momentum-resolved spectral function defined as
\begin{equation}
    \mathcal{A}\left(E,{\bf p}\right) = -\frac{1}{\pi}\mathrm{Im}\sum_l \langle {\bf p},l\rvert \mathcal{G}(E)\lvert {\bf p},l\rangle = \sum_l \langle {\bf{p}},l\rvert\delta(\mathcal{H}-E)\lvert {\bf{p}},l\rangle.
\end{equation}

\section{ \label{app:QGT}Quantum geometric tensor in real space}
The quantum geometric tensor captures the geometry of quantum states in parameter space. In crystalline systems, Bloch states $\lvert u_{n}(\mathbf{k})\rangle$ form an eigenbasis of the Hamiltonian due to translational symmetry. However, states within the same band are not orthogonal: $\langle u_{n}(\textbf{k}_1)|u_{n}(\textbf{k}_2)\rangle\neq \delta(\textbf{k}_1-\textbf{k}_2)$.
To quantify how non-orthogonal these states are, Provost and Valle~\cite{Provost1980} introduced a gauge invariant distance for isolated bands
\begin{equation}
    d^{(n)}(\textbf{k}_1,\textbf{k}_2) = \sqrt{1-|\langle u_{n}(\textbf{k}_1)|u_{n}(\textbf{k}_2) \rangle|^2}.
\end{equation}
Considering two states of the same isolated band $n$ distant from $\delta\textbf{k}$, the infintesimal interval $ds^2$ can be written as $ds^2 = \, G^{(n)}_{\mu\nu}(\textbf{k}) dk^\nu dk^\nu $, with
\begin{equation}
    G^{(n)}_{\mu\nu}(\textbf{k}) = 
    \langle \partial_\mu u_{n}(\textbf{k}) | (1- |  u_{n}(\textbf{k}) \rangle \langle u_{n}(\textbf{k})| ) |\partial_\nu u_{n}(\textbf{k}) \rangle    .
\label{QM_def}
\end{equation}
$G_{\mu\nu}(\textbf{k})$ is the \textit{quantum geometric tensor}, and $\partial_\mu=\partial/\partial k^\mu$ with $\mu,\nu = \lbrace x,y \rbrace$ in two dimensions.
Integrating the quantum geometric tensor over the Brillouin zone gives us access to the Hilbert-space volume spanned by the $n$-th state. 
The real and imaginary parts of the quantum geometric tensor are called the quantum metric and Berry curvature, respectively~\cite{bianco11,Resta2011,Abouelkomsan2023,yu2024quantumgeometryquantummaterials,verma2025quantumgeometryrevisitingelectronic}.

Despite the absence of translational symmetries of aperiodic tilings, it is possible to define a bulk, real-space version of the quantum geometric tensor as~\cite{Marrazzo2019,Abouelkomsan2023,yu2024quantumgeometryquantummaterials,verma2025quantumgeometryrevisitingelectronic}
\begin{equation}
\label{eq:QuantumGeometricTensor}
    G_{\mu\nu} 
    = \frac{2\pi}{\mathcal{A}_b}\sum_{\alpha \in \text{Bulk}}\sum_l \langle {\bf r}_\alpha,l \rvert \hat{P}\hat{{\bf r}}_{\mu}\hat{Q}\hat{{\bf r}}_{\nu}\hat{P}\lvert {\bf r}_\alpha,l\rangle,
\end{equation}
where $\mathcal{A}_b$ is the bulk area, $\hat{P}$ is the projector over occupied states, $\hat{Q}=1-\hat{P}$ is the projector onto unoccupied states and $\hat{r}_{\mu}$ is defined as
\begin{equation}
    \hat{r}_{\mu} =  \begin{cases}
        \hat{X} \quad \text{if}\quad \mu =x,\\
        \hat{Y}\quad \text{if} \quad \mu = y.
    \end{cases}
\end{equation}

\begin{figure}
    \centering
    \includegraphics[width=0.65\linewidth]{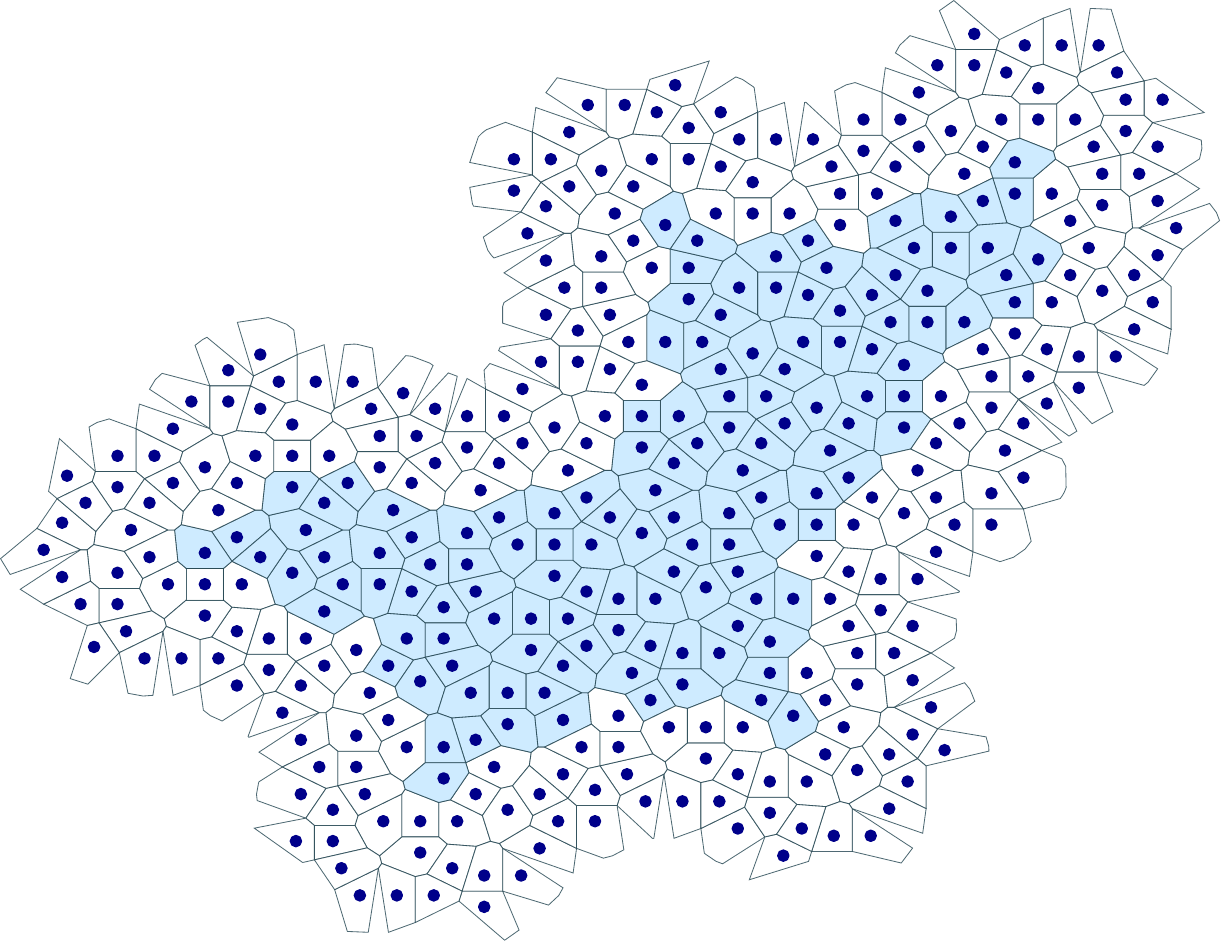}
    \caption{Voronoi tessellation of the lattice obtained for $\ell = 0.5$. The Voronoi cells corresponding to bulk sites are highlighted in light blue. The bulk area $\mathcal{A}_b$ corresponds to the sum of all the areas corresponding to individual blue cells.}
    \label{fig:Voronoi}
\end{figure}

To discuss the trace inequality~\cite{Souza2000,Parameswaran2013,Roy2014,Claassen2015,Ledwith2020,Wang2021}, as we do in the main text, it is important to normalize $G_{\mu\nu}$ consistently for all $\ell$. We achieve this by normalizing bulk-averaged quantities over a Voronoi tessellation~\cite{LejeuneDirichlet+1850+209+227,Voronoi+1908+198+287} of the system at each $\ell$, see Fig.~\ref{fig:Voronoi}.
Here, the bulk region is defined via Voronoi cells centered on lattice sites within a finite window, and the bulk area $\mathcal{A}_b$ is the sum of their individual areas.
By using the Voronoi tessellation we can conveniently account for each of the atomic site's area that contributes to $\mathcal{A}_b$. 

The quantum metric computed in the main text is then defined as the sum of the diagonal components of the quantum geometric tensor
\begin{equation}
    G = \mathrm{Re}\left[G_{xx}+ G_{yy}\right].
\end{equation}

From Eq.~\eqref{eq:QuantumGeometricTensor} one can define a local quantum metric as:
\begin{equation}
\label{eq:LocalQuantumMarker}
    g_{\bf{r}_\alpha} = 2\pi\mathrm{Re}\sum_l \langle {\bf r}_\alpha,l\lvert \hat{P}\hat{X}\hat{Q}\hat{X}\hat{P} + \hat{P}\hat{Y}\hat{Q}\hat{Y}\hat{P}\rvert {\bf r}_{\alpha},l\rangle.
\end{equation}
The quantum metric is then defined as the normalized sum over the bulk of the local quantum metric Eq.~\eqref{eq:LocalQuantumMarker}
\begin{equation}
\label{eq:avmetric}
    G = \frac{1}{\mathcal{A}_b}\sum_{\alpha \in \text{Bulk}} g_{\bf r_{\alpha}}.
\end{equation}
The quantum geometric tensor encodes important geometric and topological properties of the electronic structure: The quantum metric quantifies the spatial spread of the occupied states: $G\sim \langle \mathbf{r}^2 \rangle - \langle \mathbf{r}\rangle^2$ and is related to the localization properties of Wannier functions~\cite{Marzari1997,MarsalQuantumMetric2024}. 

Analogously, the much more commonly studied local Chern marker is defined by the antisymmetric, imaginary part of the quantum geometric tensor~\cite{bianco11,Resta2011}. Its bulk average, as defined in the main text below Eq.~\eqref{eq:LCM}, converges to the Chern number of filled states, even in systems lacking translational symmetry.

\section{Spectral localizer gap and choice of $\kappa$}
\label{app:kappa}
\begin{figure}
    \centering
    \includegraphics[width=0.8\linewidth]{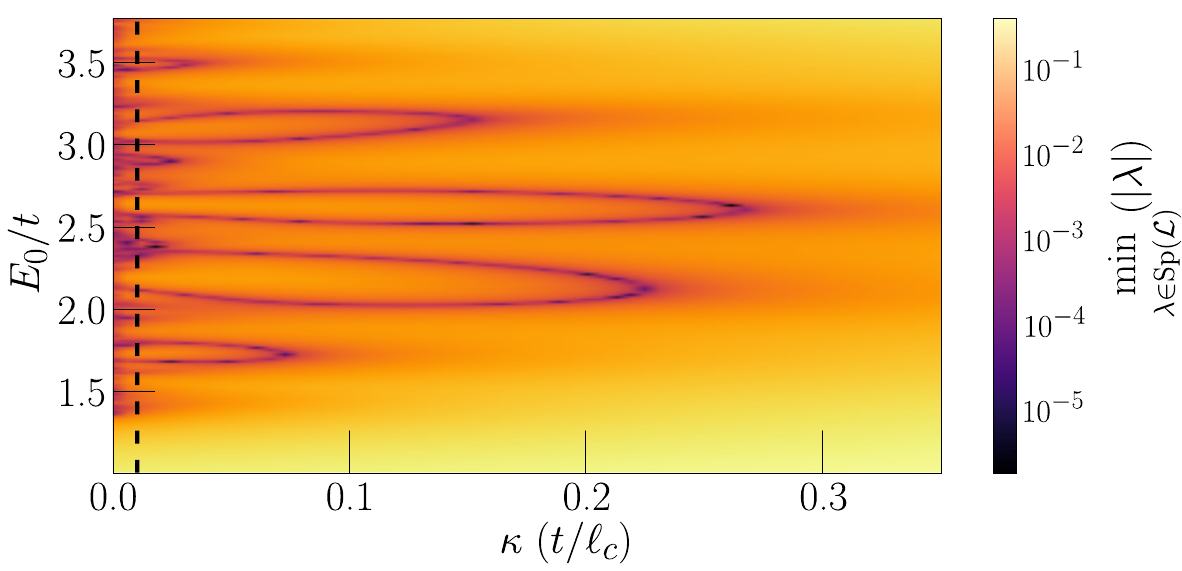}
    \caption{Localizer gap of the spectral localizer Eq.~\eqref{eq:Locapp} as a function of the reference energy $E_0/t$ and the parameter $\kappa$, measured in units of $t/\ell_c$, {using a logarithmic colormap.} Here $\ell_c=1$ is the side of the \textit{Comet} tile, see Fig.~\ref{fig:fig1}a, and sets our unit length. We compute in the bulk of the system, at a position $x_0,y_0$ chosen by computing the geometric mean of the site's coordinates, for $\ell=0.5$ and $M/t=0.5$. The vertical dashed black line corresponds to the value of $\kappa$ used in the main text. For $E_0/t = 2.65$, the energy set by $E_F$ at this $\ell$ in Fig.~\ref{fig:phase_diag}, we are within a bubble with a large localizer gap that is well defined for several orders of magnitude in $\kappa$.}
    \label{fig:KappaChoice}
\end{figure}
In this Appendix, we justify the choice of the parameter $\kappa$ used to generate Fig.~\ref{fig:phase_diag} via the spectral localizer~\cite{LORING2015,loring2019guidebottindexlocalizer,cerjan_local_2022,Cerjan_2024_Tutorial}, whose definition we repeat here for convenience
\begin{equation}
\label{eq:Locapp}
    \mathcal{L} = (\hat{\mathcal{H}}-E_0\mathbb{1}) \sigma_z + \kappa\left[(\hat{X}-x_0\mathbb{1})\sigma_x+ (\hat{Y}-y_0\mathbb{1})\sigma_y\right].
\end{equation}

To find a suitable value of $\kappa$, it is convenient to compute the localizer gap at different energies for different values of $\kappa$~\cite{Cerjan_2024_Tutorial}. The localizer gap is defined as the magnitude of the smallest eigenvalue in absolute value, $\min(\lvert\lambda\rvert)$ where $\lambda \in \mathrm{Sp}(\mathcal{L})$. For a given matrix $A$, $\mathrm{Sp}(A)$ denotes the set of all the eigenvalues of $A$ or the spectrum of $A$. 
Because the invariant can only change when the localizer gap closes, the magnitude of the localizer gap attests to the robustness of a given phase computed for a given choice of parameters $(\kappa, x_0,y_0,E_0)$. A large gap, extended in parameter space, signifies a stable phase~\cite{cerjan_local_2022,Cerjan2024crystal,Lee2025,Cerjan_2024_Tutorial}. In this sense the localizer gap acts analogously to the band-gap of crystalline systems.
In practice, for a given system size, there is not a single value of $\kappa$ but rather a window of values of $\kappa$ that capture well the topologically robust properties of the system~\cite{Liu2023localizer,Cerjan_2024_Tutorial,Lee2025}.

In Fig.~\ref{fig:KappaChoice} we plot the localizer gap as a function of both $\kappa$ and $E_0/t$ for $x_0$ and $y_0$ chosen in the bulk of the system, by computing the geometric mean of the site's coordinates. We observe the emergence of distinct bubbles where the localizer gap is large, similar to other localizer studies~\cite{cerjan_local_2022,Cerjan2024crystal,Lee2025,Cerjan_2024_Tutorial,Liu2023localizer}. These bubbles correspond to the regions in parameter space where both the Hamiltonian and the position operators meaningfully enter the calculation of the spectral localizer. 
We checked that, by choosing a value of $\kappa$ such that the energy of interest falls within one of these bubbles, the invariant obtained from the spectral localizer at that energy closely matches the bulk average of the local Chern marker at the same energy, see next section for a more detailed discussion. 

Furthermore, we observe that the bubble at $E_0/t = 2.65$, the energy set by $E_F$ at this $\ell$ in Fig.~\ref{fig:phase_diag}, is particularly extended, which ensures the robustness of the calculation with respect to different choices of the parameter $\kappa$.

\section{Comparison between the local Chern marker and the spectral localizer index \label{app:LCMvsSL}}

Formally, the spectral localizer index and the local Chern marker coincide in the small $\kappa$ limit~\cite{jezequel2025explicitequivalencespectrallocalizer}. 
Small $\kappa$ is defined with respect to the ratio between the typical energy scales of the Hamiltonian, e.g. the band width,  and the linear system size $L$.
Hence, for a given $L$ and Hamiltonian parameters, we can find a small $\kappa$ that makes the localizer index and average local Chern marker in the bulk approximately equal.
In practice, we follow the convention in the literature to fix a single $\kappa$ to calculate our phase diagrams~\cite{Cerjan_2024_Tutorial}. This is similar to the usual convention of fixing the region of integration of the local Chern marker, despite the fact that its bulk average changes depending on the spatial extent of the edge states, governed by the Hamiltonian parameters. 

\begin{figure}
    \centering
    \includegraphics[width=1.\linewidth]{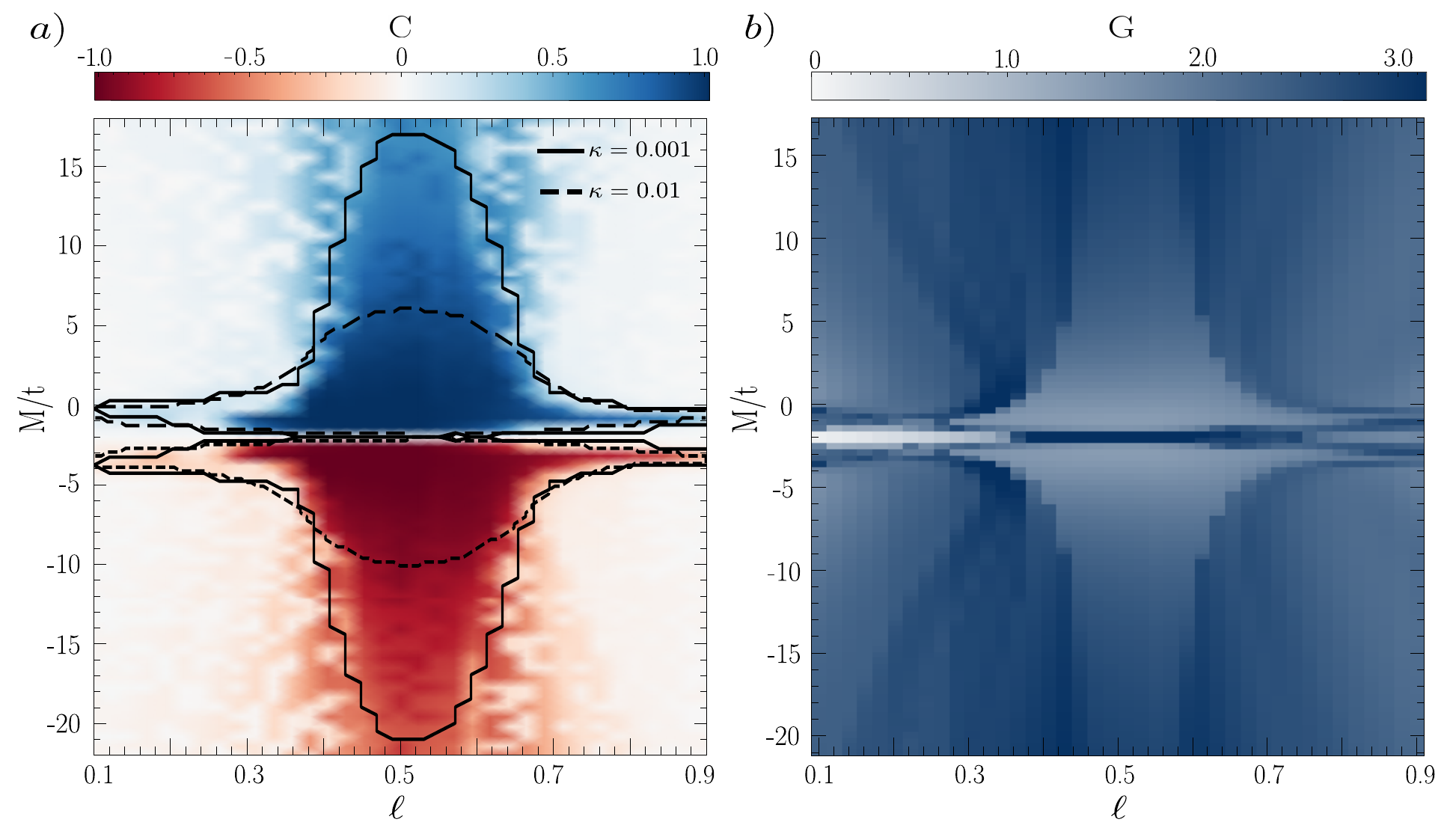}
    \caption{{Topological phase diagrams and quantum metric. (a) Phase diagram obtained from the bulk average of the local Chern marker, $C = (1/\mathcal{A}_b)\sum_{\alpha\in\text{Bulk}}C_{\textbf{r}_\alpha}$, see Eq.~\eqref{eq:LCM}, for a system of 2530 sites. The normalization area $\mathcal{A}_b$ is defined as in Fig.~\ref{fig:Voronoi}. The phase boundaries of the spectral localizer index are shown in black for $\kappa=0.01$ (dashed) and $\kappa=0.001$ (solid).
    (b) Phase diagram obtained from the bulk trace of the local quantum metric $G$ for the same system, see Eq.~\eqref{eq:avmetric}. Smaller values of $\kappa$ bring the phase diagrams computed via the spectral localizer and local Chern marker closer to agreement by capturing the topology of regions with larger quantum metric and smaller bulk gaps.}
    }
    \label{fig:phasediagramlarge}
\end{figure}

With these caveats in mind we now show that, in practice, fixing $\kappa$ for the full phase diagram recovers compatible numerical results between both methods in a large phase space, and that the agreement can be made increasingly better by tuning $\kappa$~\cite{jezequel2025explicitequivalencespectrallocalizer}. In Fig.~\ref{fig:phasediagramlarge}a-b we compare the phase diagrams of the bulk average of the local Chern marker, $C$, and the local quantum metric, $G$, and the localizer index for two different values of $\kappa$ ($\kappa=0.01, 0.001$). 
Fig.~\ref{fig:phasediagramlarge} highlights two main features. First, the phase diagrams of the localizer index and local Chern marker become increasingly similar as $\kappa$ is reduced, see Fig.~\ref{fig:phasediagramlarge}a. Second, the deviations occur at regions where the bulk average of the quantum metric is large, see Fig.~\ref{fig:phasediagramlarge}b.  
This indicates that the discrepant regions have more extended wavefunctions due to smaller spectral gaps. These properties are also reflected in the small size of the localizer gap, as seen in Fig.~\ref{fig:phase_diag}b. By reducing $\kappa$ the spectral localizer gains energy resolution, per Eq.~\eqref{eq:Locapp}, and is thus able to resolve topological properties of finer energy gaps, and the two phase diagrams become increasingly similar, as predicted by Ref.~\cite{jezequel2025explicitequivalencespectrallocalizer}. Hence, we conclude that for a given system size, smaller $\kappa$ enlarges the nontrivial regions identified by the localizer and brings the two methods into closer correspondence.

We stress that, independent of the choices above, the phase diagram shows topological regions that are larger than for the square-lattice crystalline model, as described in the main text.
Throughout the main text, we have fixed $\kappa=0.01$ for the system sizes considered. As can be seen in Fig.~\ref{fig:phasediagramlarge}a this choice is conservative.

\section{Finite-size and disorder-averaging effects}

\begin{figure}
    \centering
    \includegraphics[width=1.\linewidth]{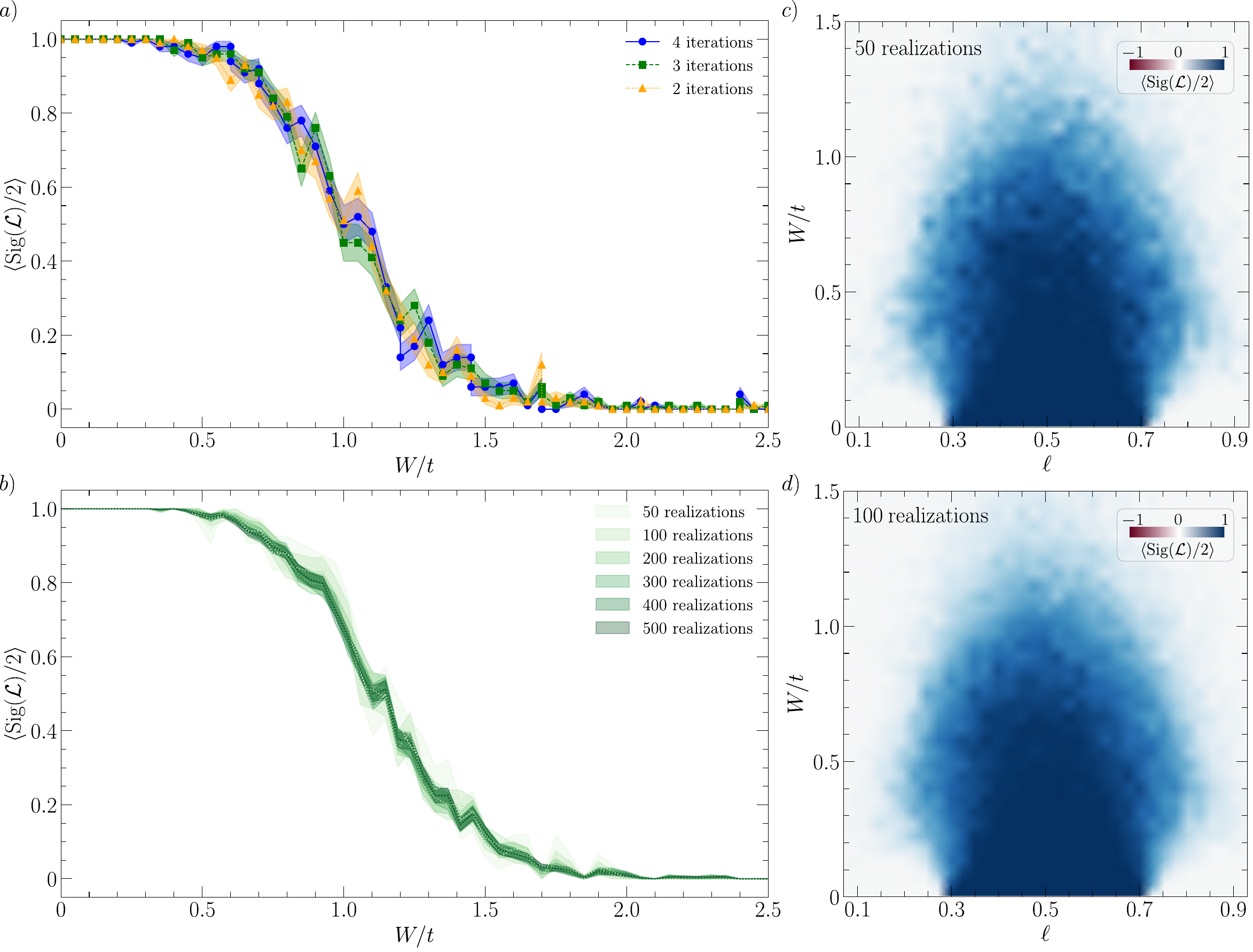}
    \caption{{Convergence of the phase diagram in the presence of Anderson disorder at 3/4 filling.
    (a) Half-signature of the spectral localizer as a function of disorder strength $W/t$ averaged over 100 disorder realizations for three different system sizes with 404, 2530 and 16496 sites, generated by two, three and four iterations of the meta-tile construction, respectively. Shaded regions indicating the standard error.
(b) Disorder-averaged half-signature at $\ell = 0.5$ for different number of disorder realizations, with standard error shown as shaded regions.
(c) and (d) show the phase diagram obtained from 50 and 100 disorder realizations, respectively.}}
    \label{fig:disorder_averaging}
\end{figure}

In this Appendix, we examine the effect of increasing the number of disorder realizations averaged on the results presented in the main text and demonstrate that our results are converged with respect to system size and disorder realizations.
Fig.~\ref{fig:disorder_averaging}a shows the half-signature at fixed parameters $\ell = 0.6$ and $M/t = -2.7$, where we vary the disorder strength $W/t$ for three system sizes: a small, medium and large system with 404, 2530 and 16496 sites generated by two, three and four iterations of the meta-tile construction, respectively. All sizes lead to a similar profile of the topological transition as a function of disorder strength.  We note that the width of the transition region, where the marker deviates from integer values, is relatively broad but comparable to those reported for a quantum spin-Hall Hamiltonian in the Ammann-Beenker tiling~\cite{Peng2021}. The model used in Ref.~\cite{Peng2021} is close to a doubled version of our Hamiltonian, with a copy per spin. Fully characterizing this transition is an interesting future research direction.

In Fig.~\ref{fig:disorder_averaging}b, we present a cut at fixed $\ell = 0.5$, averaged over different numbers of disorder realizations, ranging from 50 to 500, for a system with 2530 sites. The results converge already around 100 realizations, with only negligible differences between 100 and 500. To further confirm this convergence, Figs.~\ref{fig:disorder_averaging}c and d display the $(W/t,\ell)$ phase diagram for a system with 2530 sites, averaged over 50 and 100 disorder realizations, respectively. The two phase diagrams exhibit nearly identical phase boundaries and regions, demonstrating convergence.

\section{Example of the local quantum metric in a topological and trivial phases \label{app:metriccomp}}
\begin{figure}
    \centering
    \includegraphics[width=0.97\linewidth]{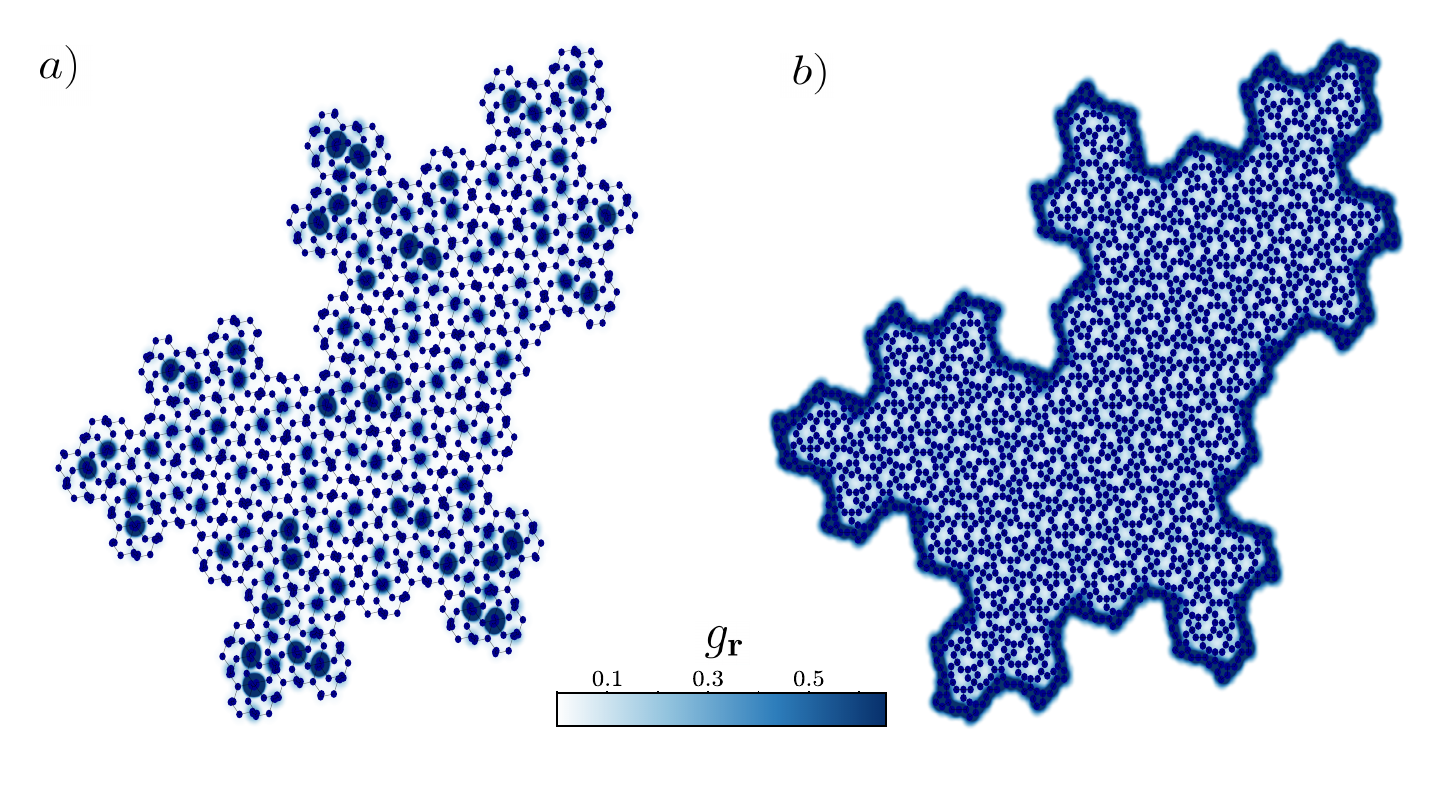}
    \caption{{Real-space distribution of the local quantum metric, $g_{\mathbf{r}}$. (a) Local quantum metric computed using Eq.~\eqref{eq:LocalQuantumMarker} in the trivial phase at 3/4 filling with $M/t = -2.7$ and $\ell = 0.1$. (b) Same calculation for $\ell = 0.5$ in the topological phase. In the trivial case (a), the metric is concentrated in localized clusters, whereas in the topological case (b), it is nearly uniform in the bulk and enhanced at the edges, reflecting the presence of a delocalized topological edge state.}}
    \label{fig:QMapp}
\end{figure}

{In this Appendix, we illustrate how the local quantum metric is spatially distributed in both trivial and topological phases. Figure~\ref{fig:QMapp} shows the quantum metric computed at $M/t = -2.7$ and 3/4 filling for a system with 2530 sites. With $\ell = 0.1$ and $\ell = 0.5$ the system lies in the trivial and topological phases, respectively. In the trivial phase, the quantum metric is localized within small clusters of sites, see Fig.~\ref{fig:QMapp}a. This is strongly indicative of localized wavefunctions, consistent with the state being in a trivial phase. By contrast, in the topological phase, the local quantum metric is nearly uniform in the bulk and reaches its maximum along the system’s edge, see Fig.~\ref{fig:QMapp}b. This is consistent with presence of a topological bulk with delocalized topological edge states.}

\end{document}